\documentclass[aps,pr,twocolumn]{revtex4}
\makeatletter

\newcommand{\Rmnum}[1]{\expandafter\@slowromancap\romannumeral #1@}
\makeatother
\usepackage{graphicx}
\usepackage{dcolumn}
\usepackage{bm}
\usepackage{CJK}
\usepackage{color}
\usepackage{amsfonts}
\usepackage{psfrag}
\usepackage{wrapfig}
\usepackage{subfigure}
\usepackage{makeidx}
\usepackage{multirow}
\usepackage{epsf}
\usepackage[colorlinks,linkcolor=red,anchorcolor=red,citecolor=blue,urlcolor=blue]{hyperref}
\usepackage{amsmath}
\usepackage{cases}
\usepackage{bm}

\begin{document}
\begin{CJK}{GBK}{song}
\title{Non-degenerate multi-rogue waves and easy ways of their excitation}
\author{Chong Liu$^{1,2,3,4}$}\email{chongliu@nwu.edu.cn}
\author{Shao-Chun Chen$^{1}$}
\author{Xiankun Yao$^{1,3}$}\email{yaoxk@nwu.edu.cn}
\author{Nail Akhmediev$^{2}$}\email{Nail.Akhmediev@anu.edu.au}
\address{$^1$School of Physics, Northwest University, Xi'an 710127, China}
\address{$^2$Department of Theoretical Physics, Research School of Physics and Engineering, The Australian National University, Canberra, ACT 2600, Australia}
\address{$^3$Shaanxi Key Laboratory for Theoretical Physics Frontiers, Xi'an 710127, China}
\address{$^4$NSFC-SPTP Peng Huanwu Center for Fundamental Theory, Xi'an 710127, China}
\begin{abstract}
In multi-component systems, several rogue waves can be simultaneously excited using  simple initial conditions in the form of a plane wave with a small amplitude single-peak perturbation. This is in drastic contrast with the case of multi-rogue waves of a single nonlinear Schr\"{o}dinger equation (or other evolution equations) that require highly specific initial conditions to be used. This possibility arises due to the higher variety of rogue waves in multi-components systems each with individual eigenvalue of the inverse scattering technique. In theory, we expand the limited class of Peregrine-type solutions to a much larger family of non-degenerate rogue waves. The results of our work may explain the increased chances of appearance of rogue waves in crossing sea states (wind generated ocean gravity waves that form nonparallel wave systems along the water surface) as well as provide new possibilities of rogue wave observation in a wide range of multi-component physical systems such as multi-component Bose-Einstein condensates, multi-component plasmas and in birefringent optical fibres.
\end{abstract}
\maketitle

\section{Introduction}
Over the past decade, the Peregrine-like waves have played a fundamental role in modeling extreme wave events \cite{RW09,Shrira10,theoryreview2017}
and attracted significant interest in various fields,
including oceanography, hydrodynamics, plasma, and optics \cite{review1,review2,review3,review4}. They are unique nonlinear excitations localised both in time and in space \cite{theoryreview2017}. Known as Peregrine waves \cite{Peregrine}, they can also appear in nonlinear superpositions \cite{JETP88,Book97,theoryreview2017}.
Such multi-rogue waves are diverse and complex structures that appear in the form of rogue wave patterns \cite{theoryreview2017}. They are
\emph{degenerate} rational solutions with several identical eigenvalues of inverse scattering technique (IST) of corresponding integrable evolution equation.
This feature distinguishes the multi-rogue waves from multi-soliton solutions. Intense theoretical and experimental studies of diverse rogue wave patterns resulted in many new findings \cite{pattern2009,pattern2012,pattern2013,He2013,He2017,pattern2020,Observation2012,Observation2016,Observation2018}. In the case of classical scalar nonlinear Schr\"odingier equation (NLSE), such degenerate solutions contain a fixed number of elementary rogue waves identical to each other when they are well separated \cite{theoryreview2017}.

Recent studies demonstrate that the degenerate multi-rogue waves exist not only in scalar nonlinear systems but also in vector ones \cite{VMRW1,VMRW2,VMRW3,VMRW4,VMRW40,VMRW41,VMRW5,VMRW6,VMRW7}. The latter involve more degrees of freedom due to the interaction between different wave components that results in more complex yet specific rogue-wave dynamics \cite{F1,F2,F3}. For example, unusual dark and four-petal rogue wave structures \cite{Z1,Z2} (which are absent in the scalar NLSE) have been found in systems governed by the vector NLSE. The existence of dark rogue wave \cite{Vobservation1} and its triplet \cite{Vobservation2} has been confirmed experimentally in fiber optics.

In this work, we discovered a new family of \emph{nondegenerate} rogue wave (RW) solutions that involve different yet well-defined IST eigenvalues of vector NLSEs.
This \emph{critical} feature means that such rogue waves are independently located in space rather than arranged in special patterns when all of them have exactly the same eigenvalue. This vital result has been missed in all previous publications  \cite{VMRW1,VMRW2,VMRW3,VMRW4,VMRW40,VMRW41,VMRW5,VMRW6}.
The actual magnitude of such eigenvalues depends on the \emph{critical} relative background wave-numbers that we define below.
Moreover, different eigenvalue for each RW means that their individual amplitude profiles also differ in contrast to a fixed Peregrine shape for all RWs in a pattern when the eigenvalues are identical.
Our non-degenerate complex solutions admit the coexistence of RWs with different amplitude profiles.
Another key point is that each of these well-separated non-degenerate RWs can be excited using simple initial conditions in the form of a plane wave with a small single-peak perturbation.

This observation has important practical consequences. Namely, if the RW solutions have different IST eigenvalues, they  can be excited simultaneously on the same background wave. Requirements to initial conditions are significantly relaxed which means that they can be excited more easily. This may result in higher chances of multiple RW excitation in crossing sea states (wind generated ocean gravity waves that form nonparallel wave systems along the water surface). The latter are described by vector evolution equations \cite{F}.
The previously known single RW excitations \cite{singlerw} turn out to be the special cases of non-degenerate RWs found here. Thus, our work significantly expands the range of RWs that may exist in nature.
\section{Model and vector RW solution}
We consider the set of $N$ vector NLSEs in dimensionless form:
\begin{equation}\label{eqmanakov}
i\frac{\partial\psi^{(j)}}{\partial t}+\frac{1}{2}\frac{\partial^2\psi^{(j)}}{\partial x^2}+ \Bigg(\sum_{j=1}^{N}|\psi^{(j)}|^2 \Bigg)\psi^{(j)}=0,~N=2,3
\end{equation}
where $\psi^{(j)}(t,x)$ are the nonlinearly coupled components of the vector wave field.
Equations (\ref{eqmanakov}) represent the Manakov model \cite{MM} when only two components are involved ($N=2$). We also consider the three-component NLSE extension ($N=3$).
These equations model the coupled nonlinear waves in variety of complex systems, ranging from the wave dynamics in crossing sea states \cite{F} and in optical fibers \cite{OF} to Bose-Einstein condensates (BEC) \cite{BEC} and even financial systems \cite{Yan}.
The physical meaning of independent variables $x$ and $t$ depends on a particular physical problem of interest. In optics, $t$ is commonly a normalised distance along the fibre while $x$ is the normalised time in a frame moving with group velocity \cite{OF}. In the case of BEC, $t$ is time while $x$ is the spatial coordinate \cite{BEC}.

The RWs under study are exact solutions of Eqs. (\ref{eqmanakov}).
They can be obtained using a Darboux transformation scheme \cite{VMRW3}:
\begin{eqnarray}
\psi^{(j)}=
\psi_{0}^{(j)}\left(1+\frac{2i(\bm{\chi}_{r}+\beta_{j})(\bm{x}+\bm{\chi}_{r}\bm{t})-2i\bm{\chi}_{i}^2\bm{t}-1}
{\gamma_{j}[(\bm{x}+\bm{\chi}_{r}\bm{t})^2+\bm{\chi}_{i}^2\bm{t}^2+1/(2\bm{\chi}_{i})^2]}\right),\label{eqrw}
\end{eqnarray}
where $\gamma_{j}=(\bm{\chi}_{r}+\beta_{j})^2+\bm{\chi}_{i}^2$, and
$\psi_{0}^{(j)}$ are the vector background plane-waves
\begin{eqnarray}
\psi_{0}^{(j)}=a_j \exp \left\{ i [{\beta_j}x + (A- \frac{1}{2}\beta_j^2)t] \right\},
\end{eqnarray}
with $A=\sum_{j=1}^{N}a_j^2$, and $a_j$, $\beta_j$ being the amplitudes and wavenumbers, respectively. Here, $\bm{x}=x-x_{01}$, $\bm{t}=t-t_{01}$, with the real parameters $x_{01}$ and $t_{01}$ describing the spatial and temporal centers of the RW.
The values, $\bm{\chi}_r\equiv\textrm{Re}[\bm{\chi}]$ and $\bm{\chi}_i\equiv\textrm{Im}[\bm{\chi}]$ are the real and imaginary parts of the eigenvalue of the Lax pair $\bm{\chi}$
associated with Eqs. (\ref{eqmanakov}). It plays a vital role in the wave structure and the possibility of RW formation.

The wavenumbers must satisfy the constraints:
$$
{\beta_1}\neq{\beta_2}\neq{\beta_3}.
$$
Here, we set
\begin{eqnarray}
{\beta_1}=-{\beta_2}=\beta,~~~ \textrm{for}~~~ N=2,\\
{\beta_1}=-{\beta_3}=\beta,~{\beta_2}=0,~~~ \textrm{for} ~~~N=3.
\end{eqnarray}
Only the relative wavenumber $\beta$ is important since it cannot be eliminated through a Galilean transformation. We can also equalise the backgrounds: $a_j=a$. The latter condition is not essential but simplifies the presentation reducing the number of free parameters in the illustrative material.

\section{eigenvalue analysis and Non-degenerate multi-rogue waves}
For the $N$-component NLSEs, the eigenvalue must satisfy the relation
\begin{eqnarray}
1+\sum_{j=1}^{N}\frac{a^2}{(\bm{\chi}+\beta_j)^2}=0.\label{eqchi}
\end{eqnarray}
In particular, for $N=2$, the explicit expressions are given by,
\begin{eqnarray}
\bm{\chi}^2_{(1)}&=&\beta^2-a^2+ \sqrt{a^4-4a^2\beta^2},\label{eqchi-2-1}\\
\bm{\chi}^2_{(2)}&=&\beta^2-a^2- \sqrt{a^4-4a^2\beta^2},\label{eqchi-2-2}
\end{eqnarray}
while for $N=3$,
\begin{eqnarray}
3\bm{\chi}^2_{(1)}&=&2 \beta ^2-3a^2+\frac{\mathcal{A}}{\mathcal{B}}+\mathcal{B},\label{eqchi-3-1}\\
3\bm{\chi}^2_{(2)}&=&2 \beta ^2-3a^2-\frac{2}{1-i \sqrt{3}}\frac{\mathcal{A}}{\mathcal{B}}-\frac{1-i \sqrt{3}}{2}\mathcal{B},\label{eqchi-3-2}\\
3\bm{\chi}^2_{(3)}&=&2 \beta ^2-3a^2-\frac{2}{1+i \sqrt{3}}\frac{\mathcal{A}}{\mathcal{B}}-\frac{1+i \sqrt{3}}{2}\mathcal{B}.\label{eqchi-3-3}
\end{eqnarray}
with $$\mathcal{A}=\beta ^4-12 \beta ^2 a^2 +9a^4,$$ and
\begin{eqnarray}
&\mathcal{B}^3=54 a^4\beta ^2-\beta ^6-36 a^2\beta ^4-27a^6\nonumber\\
&+3 \sqrt{3} a\beta ^2\sqrt{4 \beta ^6+27 a^2(\beta ^2-a^2)^2}.\nonumber
\end{eqnarray}
Here, $(-1)^{1/3}=(1+i \sqrt{3})/2$ for $\mathcal{B}^3<0$.

We can see from (\ref{eqrw}) that the sign of $\bm{\chi}_i$ has no effect on the solution while the sign of $\bm{\chi}_r$ determines the `velocity' ($x+\bm{\chi}_{r}t$) and the `phase' [$2i(\bm{\chi}_{r}+\beta_{j})$] of the RWs.

\emph{Another important point is that although Eq. (\ref{eqrw}) is a valid solution of Eq. (\ref{eqmanakov}) for every $\bm{\chi}$, not all solutions are realistic.} For example,
when $\beta=0$, the system (\ref{eqmanakov}) consists of $N$ identical NLSE components and the corresponding eigenvalues must satisfy the relation \begin{eqnarray}
\bm{\chi}^2=-Na^2.
\end{eqnarray}
In this particular case, only Eqs. (\ref{eqchi-2-2}) and (\ref{eqchi-3-2}) define the valid eigenvalues for $N=2$ and $N=3$, respectively.
The eigenvalues defined by Eqs. (\ref{eqchi-2-1}), (\ref{eqchi-3-1}), and (\ref{eqchi-3-3}) vanish because $\beta=0$.

Moreover, there are relative wavenumbers $\beta=\beta_c$ (we call them \emph{critical})  that determine the validity of the eigenvalues. They can be found from Eqs. (\ref{eqchi-2-1})-(\ref{eqchi-3-3}) by setting $\bm{\chi}^2_{(j)i}=\textmd{Im}[\bm{\chi}^2]=0$:
\begin{eqnarray}
\beta_c=\pm a/2,~~ \textrm{for}~~N=2, \label{eqcwb-2}\\
\beta_c=\pm a\sqrt{6-3\sqrt{3}},~~ \textrm{for}~~N=3.\label{eqcwb-3}
\end{eqnarray}
Now, the eigenvalues (\ref{eqchi-2-1})-(\ref{eqchi-3-3}) are valid only in the regions of $|\beta|\geq|\beta_c|$. In the opposite case, $|\beta|<|\beta_c|$, only the eigenvalues defined by Eqs. (\ref{eqchi-2-2}) and (\ref{eqchi-3-2}) are valid for $N=2$ and $N=3$, respectively.

\begin{figure}[htb]
\centering
\includegraphics[width=85mm]{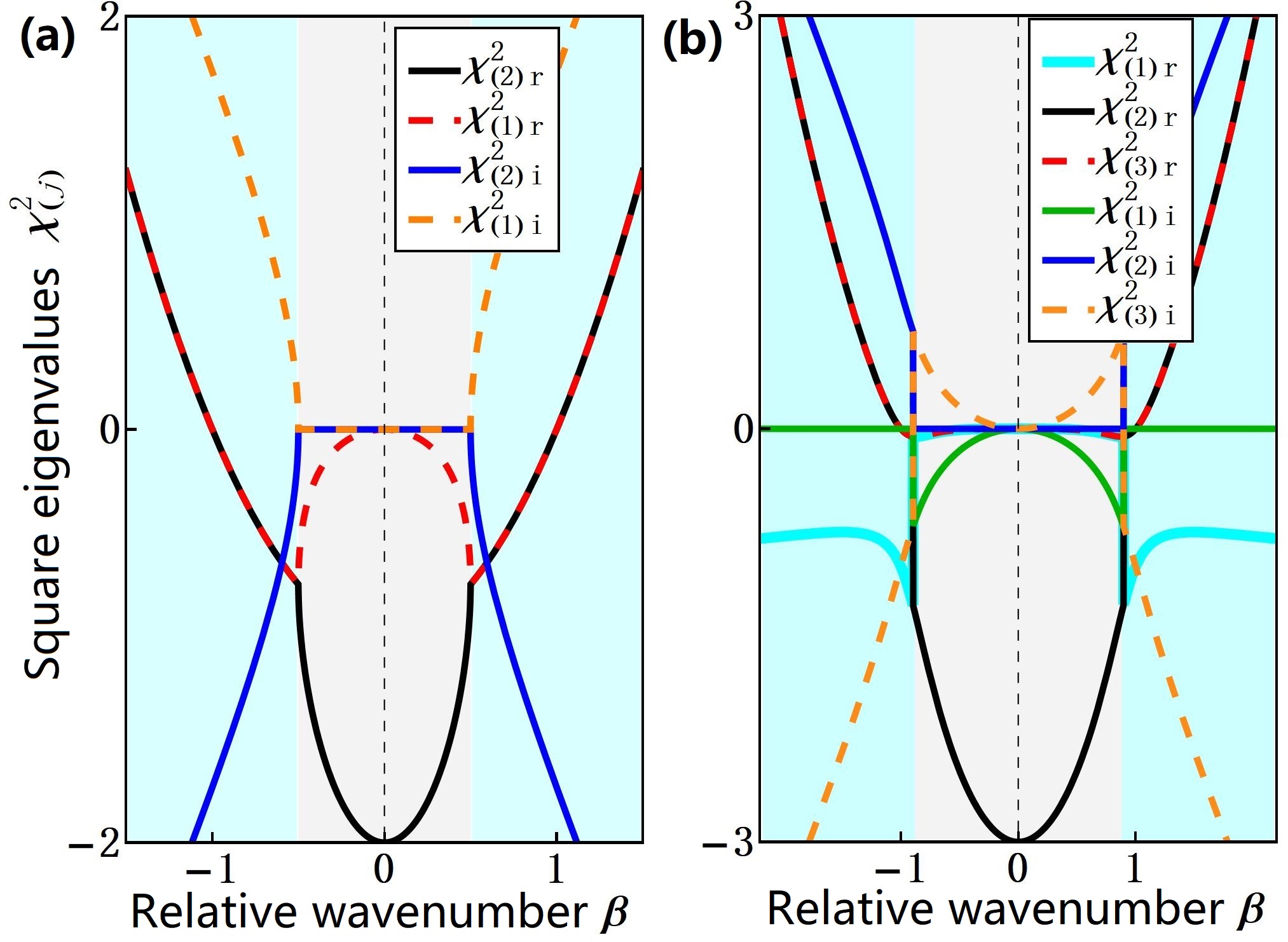}
\caption{Real and imaginary parts of the square eigenvalues $\bm{\chi}_{(j)}^2$, Eqs. (\ref{eqchi-2-1})-(\ref{eqchi-3-3}), versus $\beta$ for (a) $N=2$, and (b) $N=3$.
The grey areas correspond to degenerate ($|\beta|\leq|\beta_c|$) while the cyan areas to non-degenerate ($|\beta|>|\beta_c|$) regimes.
The background $a=1$. }  \label{f2-1}
\end{figure}

The role of $|\beta_c|$ is illustrated in Fig. \ref{f2-1}. This figure
 shows the real and imaginary parts of the squared eigenvalues $\bm{\chi}^2_{(j)}$, given by Eqs. (\ref{eqchi-2-1})-(\ref{eqchi-3-3}), versus $\beta$. In each case, there are two areas separated by $|\beta_c|$ where the physics is completely different.
In the grey areas $|\beta|\leq|\beta_c|$, the system (\ref{eqmanakov}) reduces to the scalar NLSE when $|\beta|\rightarrow0$. In this limit, the RW in each component is transformed into the standard Peregrine RW. Only Eqs. (\ref{eqchi-2-2}) and (\ref{eqchi-3-2}) are valid in this case. Real parts of the squared eigenvalues are negative
 and $\bm{\chi}_{(2)}=\pm i\bm{\chi}_i$. As shown in Fig. \ref{f2-1}, this limit is described by the black solid curves that end at the points
 $$(\beta, \bm{\chi}^2_{(2)})=(0,-Na^2).$$
 This limit leads to decoupling of Eqs. (\ref{eqmanakov}).

The areas $|\beta|\leq|\beta_c|$ are the regions of simple vector generalisation of the scalar RWs of a single NLSE as there is a one-to-one correspondence between $\beta$ and $\bm{\chi}_{(2)}$. The sign of $\bm{\chi}_i$ has no effect on the RW solution.
This situation admits only one type of RW
(shown in Fig. \ref{f1}(a) below).
On the contrary, in the areas $|\beta|>|\beta_c|$, shown in cyan colour all eigenvalues are valid. Specifically, the square eigenvalues (\ref{eqchi-2-1}) and (\ref{eqchi-2-2}) in Fig. \ref{f2-1}(a) ($N=2$) have the same (nonzero) real part but have imaginary parts of opposite sign. As a consequence, $$\bm{\chi}_{(1)}=\pm(i\bm{\chi}_i+\bm{\chi}_r)$$ and $$\bm{\chi}_{(2)}=\pm(i\bm{\chi}_i-\bm{\chi}_r).$$ Thus, for any $\beta$ in the region of $|\beta|>|\beta_c|$, there are two valid eigenvalues with real parts of opposite sign. This implies the coexistence of two non-degenerate RWs  (shown in Figs. \ref{f1}(b) and (c) below).

Similar conclusions can be achieved in the case $N=3$. The square eigenvalues  (\ref{eqchi-3-2}) and (\ref{eqchi-3-3}) for this case are shown in Fig. \ref{f2-1}(b).
However, for $N=3$, there is one more squared eigenvalue with negative real part given by Eq. (\ref{eqchi-3-1}). This leads to the coexistence of three non-degenerate RWs for $N=3$ (shown in Figs. \ref{f2}(b) and \ref{f2}(c) below).

Further analysis shows that the RWs (\ref{eqrw}) admit symmetries with respect to the sign change of $\beta$, $\bm{\chi}$, and $x$:
\begin{eqnarray}
\label{eqs1}
&\psi^{(N)}(\beta)=\psi^{(1)}(-\beta),\\
&\psi^{(j)}\left(\bm{\chi}_{i}\right)=\psi^{(j)}\left(-\bm{\chi}_{i}\right),\\
&\psi^{(j)}\left(x,\bm{\chi}_{r}\right)=\psi^{(N+1-j)}\left(-x,-\bm{\chi}_{r}\right).\label{eqs3}
\end{eqnarray}
The symmetries (\ref{eqs1})-(\ref{eqs3})  provide deeper understanding of the non-degenerate RW phenomena revealed here.

\begin{figure}[htb]
\centering
\includegraphics[width=85mm]{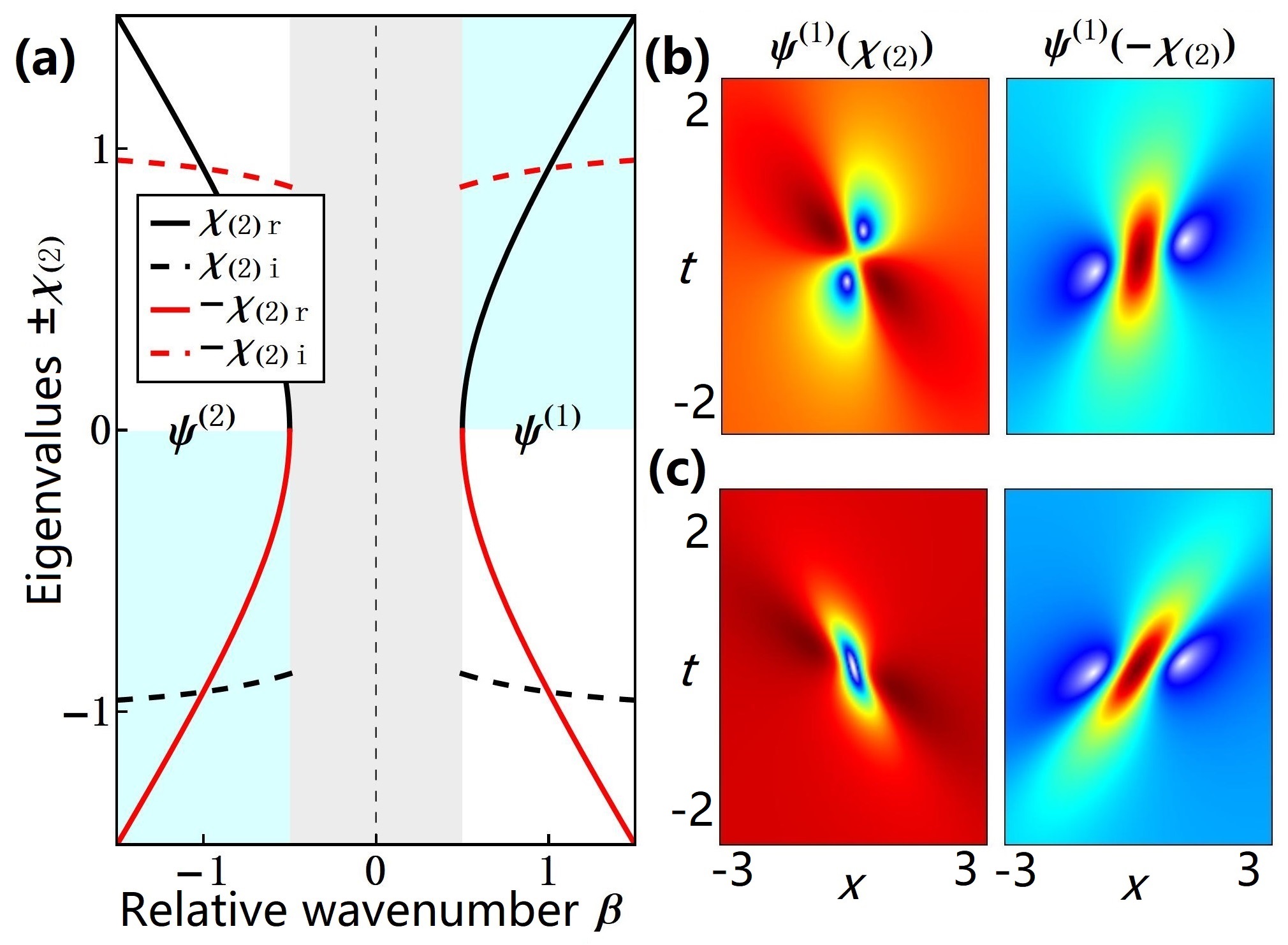}
\caption{
(a) Real and imaginary parts of the eigenvalues $\pm\bm{\chi}_{(2)}$ vs $\beta$ for $N=2$.
The grey area [$|\beta|\leq|\beta_c|$, Eq. (\ref{eqcwb-2})] corresponds to the degenerate regime.
The regions $|\beta|>|\beta_c|$ produce $N$ different types of non-degenerate RWs.
The cyan areas correspond to four-petal or dark RWs, while the white areas correspond to the Peregrine-like RWs.
(b) RW profiles $|\psi^{(1)}(x,t)|$ [given by Eqs. (\ref{eqrw})] in the non-degenerate regime for $\beta=0.6$.
(c) Same as in (b) but for $\beta=0.9$.
Here, $a=1$, $x_{01}=t_{01}=0$.}  \label{fn2}
\end{figure}

Let us first concentrate on the case of $N=2$. Fig. \ref{fn2} shows the eigenvalues $\pm\bm{\chi}_{(2)}$ given by Eq. (\ref{eqchi-2-2}) and the corresponding amplitude profiles of RWs. We only show the $\psi^{(1)}$ component of RWs as the other one is related to $\psi^{(1)}$ through the symmetries (\ref{eqs1})-(\ref{eqs3}). As explained above, the RWs with opposite signs of the real parts of eigenvalues [$\pm\bm{\chi}_{(2)r}\neq0$] in the non-degenerate regime ($|\beta|>|\beta_c|$)  have different wave profiles.
Left hand side panels in Figs. \ref{fn2}(b) and \ref{fn2}(c) show $\psi^{(1)}(\bm{\chi}_{(2)})$.
Namely, Fig. \ref{fn2}(b) shows a four-petal wave profile while Fig. \ref{fn2}(c) shows a dark RW structure. On the other hand, $\psi^{(1)}(-\bm{\chi}_{(2)})$ shown on the right hand side panels
always has a Peregrine-like waveform. These results show that the RWs can be a combination of a Peregrine-like and a four-petal or a dark structure.

\begin{figure}[htb]
\centering
\includegraphics[width=85mm]{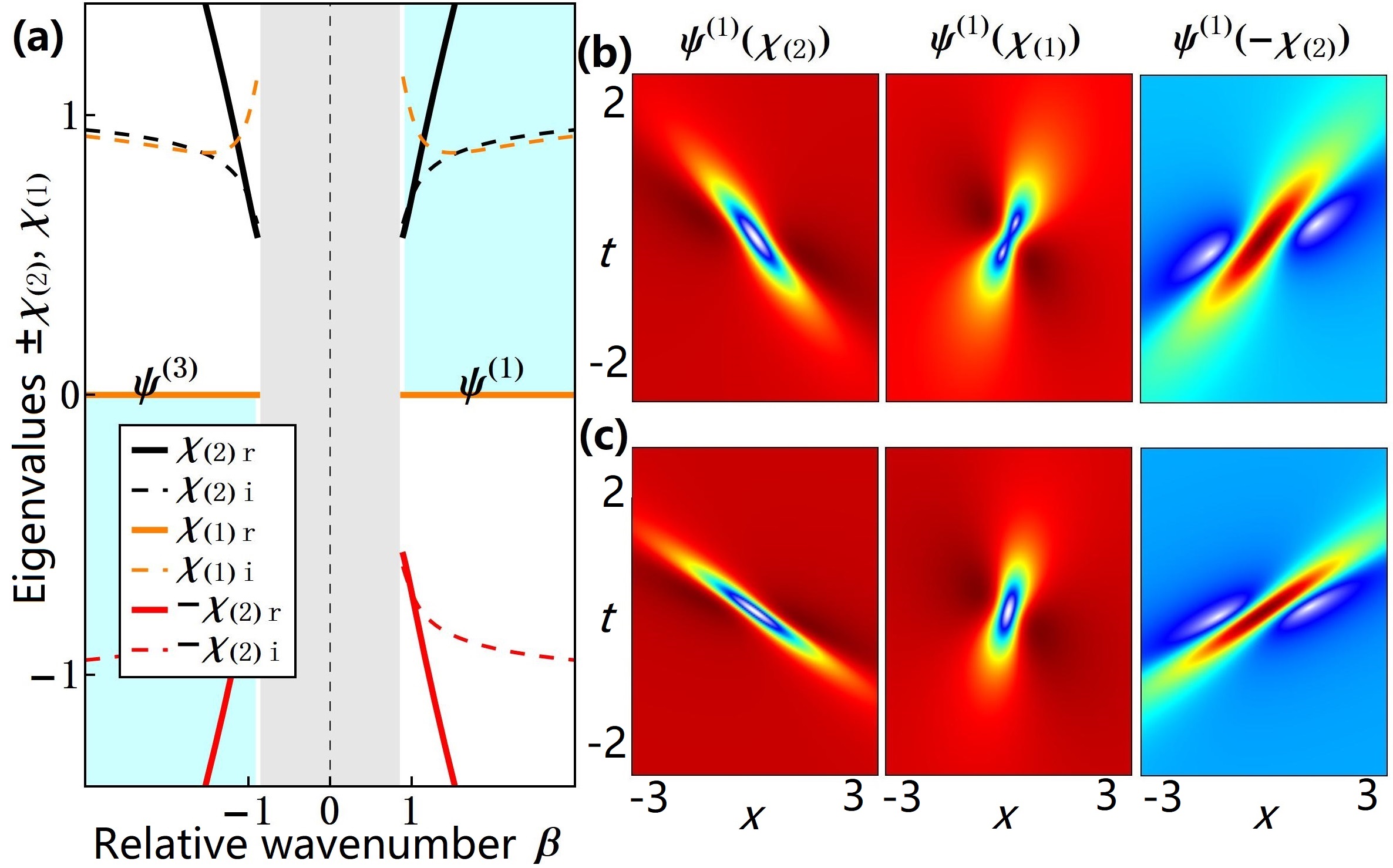}
\caption{Same as in Fig. \ref{fn2} but for $N=3$. (a) Real and imaginary parts of the eigenvalues [$\pm\bm{\chi}_{(2)}$, $\bm{\chi}_{(1)}$] vs $\beta$. Non-degenerate RW profiles $|\psi^{(1)}(x,t)|$ for (b) $\beta=1.3$ and (c) $\beta=2$.
} \label{fn3}
\end{figure}

For $N=3$, there are three non-degenerate eigenvalues [$\bm{\chi}_{(1)}$, $\pm\bm{\chi}_{(2)}$] shown in Fig. \ref{fn3}(a). The corresponding wave profiles $|\psi^{(1)}|$ are shown in Figs. \ref{fn3}(b-c). The three wave profiles are all different. The profiles $|\psi^{(1)}(\bm{\chi}_{(2)})|$ and $|\psi^{(1)}(-\bm{\chi}_{(2)})|$ shown at the left and right hand side panels of Figs. \ref{fn3}(b-c) can be described as the dark and the Peregrine-like modes, respectively. The profile $|\psi^{(1)}(\bm{\chi}_{(1)})|$ shown at the central panels of Figs. \ref{fn3}(b-c) can be described as the four-petal (b) or the dark (c) structure.

So far, we studied only the individual RWs. Multi-RW solutions can be obtained using the higher-order iterations of the Darboux transformation scheme \cite{method,Erkintalo}. Calculations are cumbersome but the idea is simple \cite{Erkintalo}. Each new step in these calculations adds one more RW to the existing pattern. The eigenvalues $\bm{\chi}_{(j)}$ chosen at each iteration $m$ ($m=1, ..., M$) define the RW profile while the parameters $(x_{0m},t_{0m})$ define its location.  The resulting $M$th-order solution contains $M$ elementary RWs, each associated with the given eigenvalue $\bm{\chi}_{(j)}$ and the coordinates of its centre $(x_{0m},t_{0m})$. Since the $N$ vector NLSEs have $N$ different valid eigenvalues in the non-degenerate case, we have $2\leq M\leq N$.
The details are presented in Appendix.
Our approach is different from the degenerate multi-RW case where the eigenvalues coincide.
The spatiotemporal distribution of such non-degenerate RWs strongly depends on the relative separations in both $x$ and $t$, i.e., $\delta x=\{x_{01}, ..., x_{0M}\}$, and $\delta t=\{t_{01}, ..., t_{0M}\}$.


\begin{figure}[htb]
\centering
\includegraphics[width=85mm]{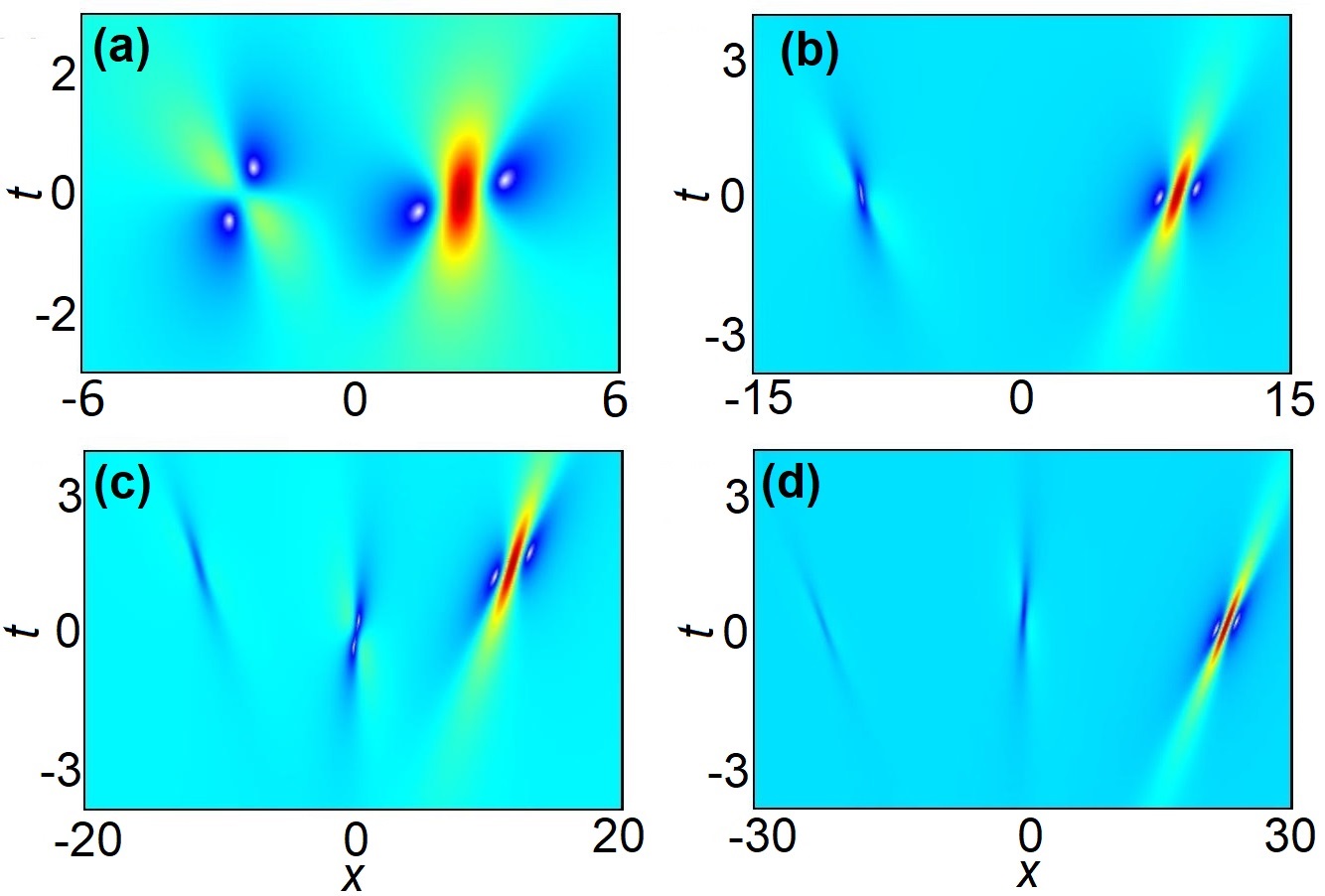}
\caption{Wave profiles $\psi^{(1)}$ of higher-order non-degenerate RWs
for (a,b) $N=2$ and (c,d) for $N=3$. The values of
parameters are (a) $\beta=0.6$, $\delta x =\{-1.1, 2.3\}$, $\delta t =\{-0.2, 2.5\}$; (b) $\beta=0.9$, $\delta x =\{-8.727, 8.727\}$, $\delta t =\{0.7, -1.45\}$; (c) $\beta=1.3$, $\delta x =\{1.78, -7.61, 15.21\}$, $\delta t =\{0.1, -0.3, 3.6\}$; (d) $\beta=2$, $\delta x =\{1.7, -18, -26.21\}$, $\delta t =\{-0.2, -1.2, 1.5\}$.
}\label{f3}
\end{figure}

Figure \ref{f3} shows four examples of such higher-order solutions.
Only $\psi^{(1)}$ component is shown. Figures \ref{f3}(a-b) corresponds to the case of $N=2$
while Figures \ref{f3}(c-d) correspond to the case $N=3$.
This figure shows that $N$ RWs do coexist on the same background wave.
The values of $\beta$ and the relative locations $\delta t$ and $\delta x$ of RWs [differences of coordinates $(x_{0m},t_{0m})$] are shown in the figure caption.

\section{Easy ways of their excitation}

Having exact solutions at hand, it is easy to excite the RWs in numerical simulations. Taking the solution (\ref{eqrw}) at sufficiently large negative $t=t_0$ as the initial condition, we obtain exactly the same profiles as given by Eq. (\ref{eqrw}).
This initial condition $\psi^{(j)}=\psi^{(j)}(t=t_0,x)$ is a constant background with a localised perturbation of a specific shape defined by the eigenvalue $\chi$ and the value of $t_0$. Numerical modelling does follow the exact solutions shown in Figs. \ref{fn2}(b,c) and \ref{fn3}(b,c) as we have checked in our simulations.

Our next task is to demonstrate that the above RWs can be excited with a wider class of initial conditions than the one described above.
In order to do that, we have solved Eqs. (\ref{eqmanakov}) numerically using the split-step Fourier method \cite{LA2021,Liu2021}.
As Eqs. (\ref{eqmanakov}) are integrable, in principle, the initial value problem can also be solved exactly. However, for a nonzero background, this problem is highly complex even in the case of a single NLSE. Thus, here, we limit ourselves with numerical examples based on the initial conditions in the form of a single-peak
perturbation on top of plane waves:
\begin{equation}\label{eqic}
\psi^{(j)}=\psi_{0}^{(j)}\left[1+\epsilon L_p(x/x_0)\right],
\end{equation}
where the single-peak perturbation $L_p(x/x_0)$ is either a sech-function $L_p=\textmd{sech}(x/x_0)$ or a Gaussian $L_p=\exp{(-x^2/x_0^2)}$.

The background plane waves $\psi_{0}^{(j)}$ in (\ref{eqic}) are the same as in the exact solution (\ref{eqrw}).
As it follows from the analytic results above, there is a threshold value of $\beta$ separating qualitatively different rogue wave patterns. Thus, we choose $\beta$ values in numerical simulations at different sides of this threshold. In each case, $x_0$ is the width of the perturbation. Its value in simulations should be comparable with the width of the RW. We used the range of values $x_0\in[3,12]$. The amplitude of the perturbation $\epsilon$ is a small real parameter. Its actual value is not critical once $\epsilon \ll1$.

\begin{figure}[htb]
\centering
\includegraphics[width=85mm]{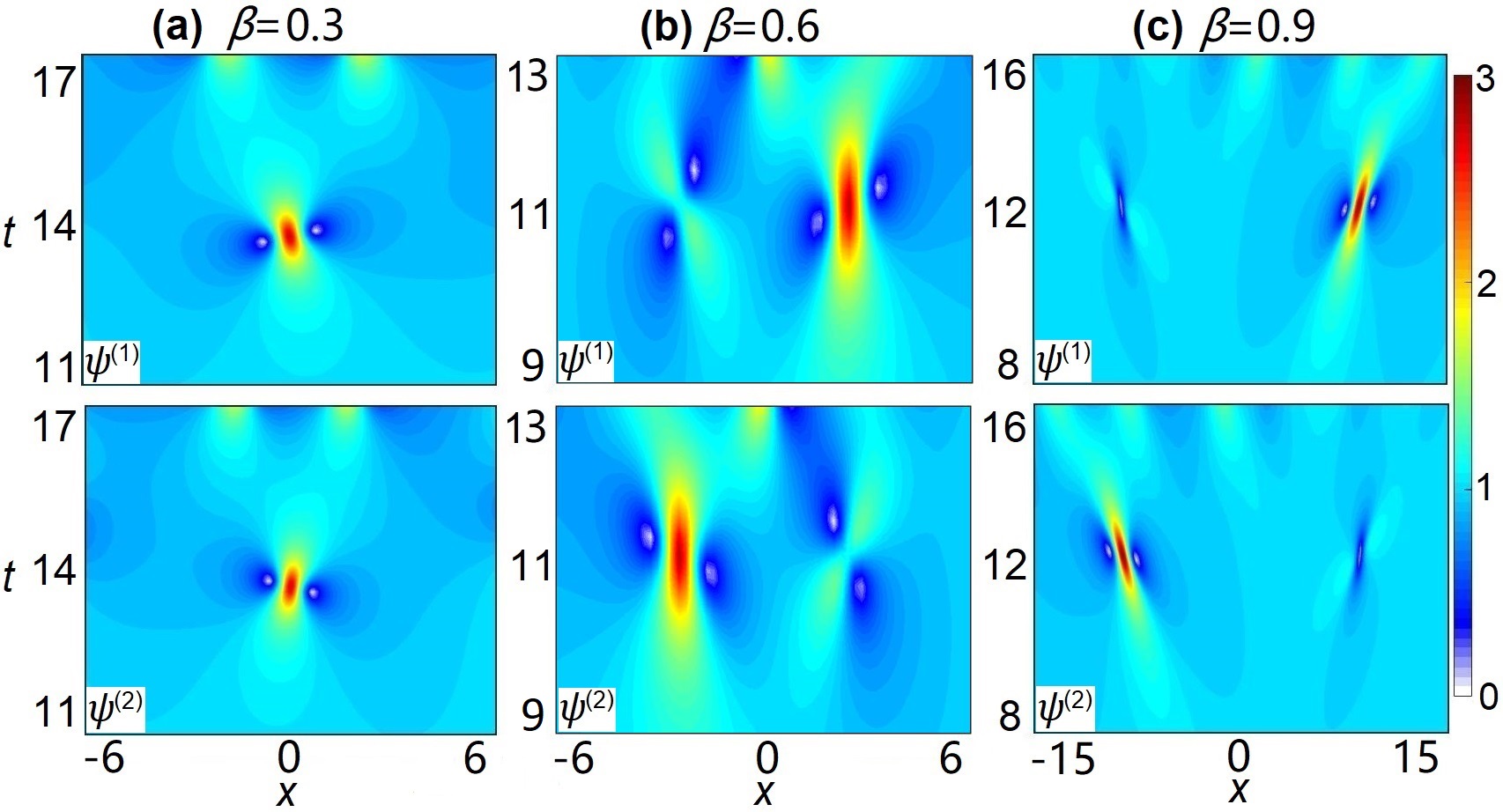}
\caption{Two-component vector RWs ($|\psi^{(j)}(x,t)|$), excited
using initial conditions (\ref{eqic}) with a Gaussian perturbation.
The three cases correspond to different $\beta$:
(a) $\beta=0.3$, (b) $\beta=0.6$, and (c) $\beta=0.9$. Others are $\epsilon=0.01$, $a=1$, $x_0=6$.
} \label{f1}
\end{figure}

\begin{figure}[htb]
\centering
\includegraphics[width=85mm]{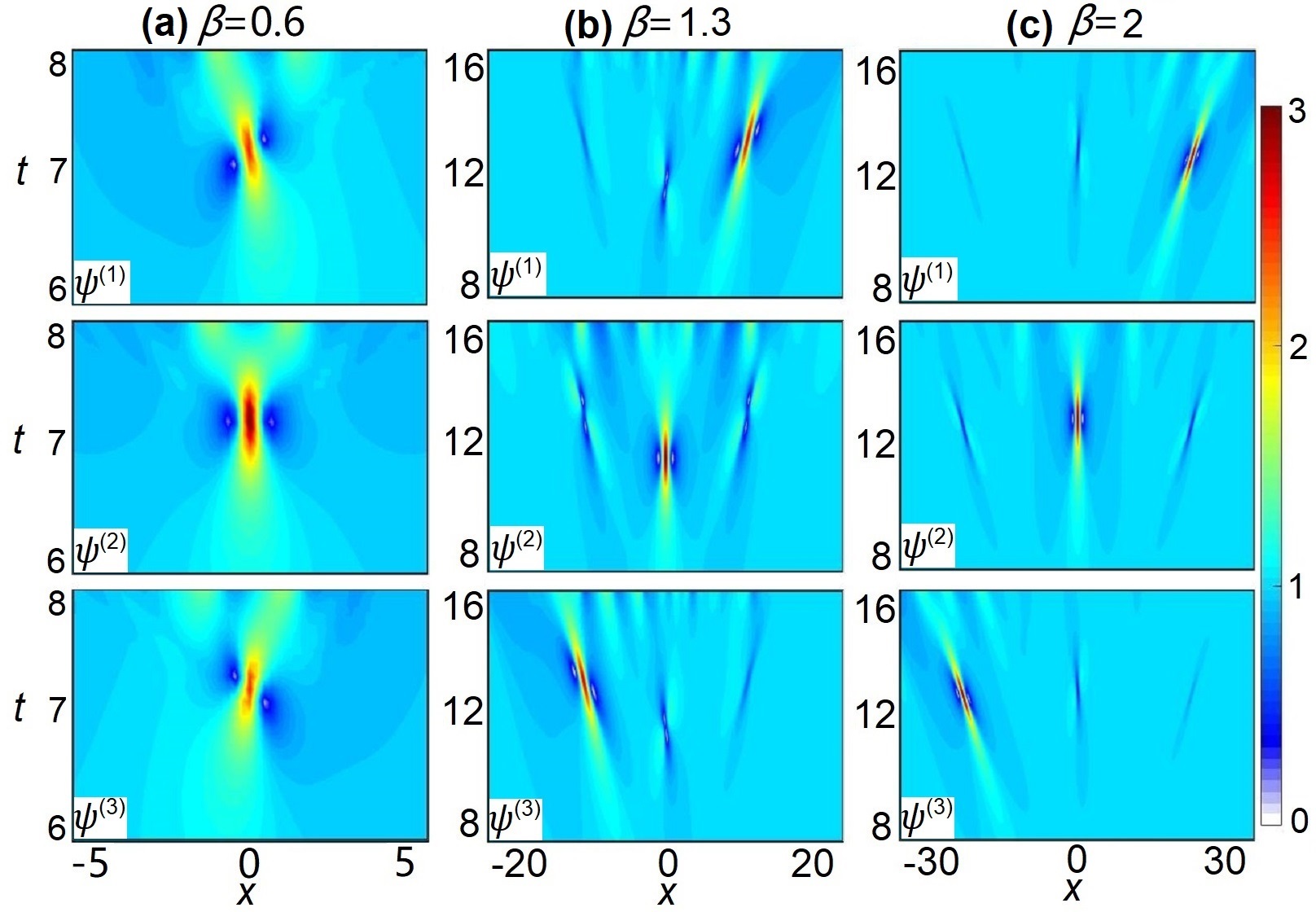}
\caption{Three-component vector RWs, $|\psi^{(j)}(x,t)|$,
excited using initial conditions (\ref{eqic}) with a Gaussian perturbation. The three cases correspond to different $\beta$:
(a) $\beta=0.6$, (b) $\beta=1.3$, and (c) $\beta=2$. Others are the same as in Fig.\ref{f1}.
} \label{f2}
\end{figure}

Figure \ref{f1} shows the results of numerical simulations started from the initial conditions (\ref{eqic}) for the case $N=2$ while Fig. \ref{f2} shows the results for $N=3$. In each case, the RWs predicted analytically are excited.
As mentioned, the major factor that influences the excitation of RWs is the wavenumber $\beta$. It increases from the left to the right panels in each figure. When the value of $\beta$ is below critical as in Figs. \ref{f1}(a) and \ref{f2}(a), a single Peregrine-like RW is excited in every component of the coupled wave field in accordance with the theoretical predictions.
When the value of $\beta$ is above critical, $N$ well-separated RWs are excited. Namely, two RWs can be seen in Figs. \ref{f1}(b) and \ref{f1}(c) and three RWs in Figs. \ref{f2}(b) and \ref{f2}(c).
Their maxima are located at different components of the coupled wave field.

Figure \ref{f1}(b) calculated for $\beta=0.6$ shows the excitation of two RWs at around $t\approx11$ similar to the RW in Fig. \ref{fn2}(b). They are symmetrically located to the left and right from the centre line $x=0$ and their maxima in one of the components (either $\psi^{(1)}$ or $\psi^{(2)}$) is much higher than in the other component.
Figure \ref{f1}(c) calculated for $\beta=0.9$ shows the excitation of two RWs (at $t\approx12.5$)
similar to the RWs in Fig. \ref{fn2}(c).
The separation of the two RWs here is larger than in Fig.\ref{f1}(b). Note the different scale of the two plots along the $x$-axis.
The maximal amplitudes of the two RWs are also reversed in $\psi^{(1)}$ and $\psi^{(2)}$ components.

When $N=3$, the critical value of $\beta$ is given by Eq. (\ref{eqcwb-3}). There are three RWs produced when $\beta>\beta_c$.
These results are shown in Figs. \ref{f2}(b) and \ref{f2}(c).
For $\beta=1.3$, the RWs in numerical simulations are similar to the RWs in Fig. \ref{fn3}(b)
while for $\beta= 2$ they are similar to the RWs in Fig. \ref{fn3}(c) except for
the RW maxima that each appears in different field component.

A small noise that accumulates in numerical simulations of this type of equations \cite{Ablowitz} does not influence the RW excitation. However, it leads to modulation instability of the background and chaotic evolution that occurs far beyond the $t$-range shown in the above figures.

The patterns of non-degenerate RWs shown in Figs. \ref{f1}(b)(c), \ref{f2}(b)(c) coincide with the higher-order RW solutions with an appropriate choice of the relative separations shown in Fig. \ref{f3}. Thus, the numerical simulations above confirm the discovery of non-degenerate RWs.


\section{Conclusions}
Our results are important in practice since they reveal the fact that in multicomponent systems, the degeneracy of RWs can be lifted and each RW can be assigned its individual eigenvalue. Clearly, variety of RW profiles also is large in comparison with a single NLSE case which is fixed to the Peregrine wave. Moreover, RWs and their combinations in multicomponent systems can be easily excited using initial conditions in the form of a plane wave with a small single peak perturbation. In contrast, RW excitation in a single NLSE case requires accurately prepared wave profile that leads to the excitation of a Peregrine wave. Excitation of multi-RWs then is increasingly difficult. In a non-degenerate case, presented here, several RWs can be easily excited with a simple initial condition described above. As the oceanic waves in the crossing sea states are multicomponent systems \cite{F}, we would expect that according to our theory, the enhanced probability of the appearance of RWs can be expected due to the simplicity of their excitation.

Considering the wide range of applications of vector NLSEs, our analysis predicts simpler ways to excite the non-degenerate RWs in a variety of vector nonlinear systems. Patterns of RWs have been found for multiplicity of evolution equations. We expect that non-degenerate RWs do exist for other coupled evolution equations that admit Peregrine-type (degenerate) RWs.

\section*{ACKNOWLEDGEMENTS}
This work is supported by the NSFC (Grants No. 12175178, No. 12004309, No. 12022513, and No. 12047502), the Major Basic Research Program of Natural Science of Shaanxi Province (Nos. 2017KCT-12).


\begin{appendix}
\section{Exact solutions of nondegenerate multi-RWs}
The exact $M$th-order solutions describing non-degenerate multi-RWs of the set of $N$ vector NLSEs are constructed by the Darboux transformation. 
The equivalent linear system (Lax pair) of vector NLSEs (\ref{eqmanakov}) is given by
\begin{eqnarray}
\Psi_x=U\Psi,~~\Psi_t=V\Psi,\label{lax}
\end{eqnarray}
with
\begin{eqnarray}
U=i\left[\frac{\lambda}{2}(S+I_{N+1})+Q\right],~~~~~~~~~~~~~~~~\nonumber\\
V=i\left[\frac{\lambda^2}{4}(S+I_{N+1})+\frac{\lambda}{2}Q-\frac{1}{2}S(Q^2+iQ_x)+B\right],\nonumber
\end{eqnarray}
where
\begin{eqnarray}
Q=\begin{pmatrix} 0 & {\psi_{0}^\dag}  \\ {\psi_{0}} & 0 \end{pmatrix},
~~S=\textmd{diag}\{1,-I_N\}.
\nonumber
\end{eqnarray}
Here, ${\psi_{0}}=(\psi_{0}^{(1)}, \psi_{0}^{(2)},...,\psi_{0}^{(N)})$ and $\dag$ represents the Hermite conjugation.
Moreover, $\lambda$ denotes the spectral parameter, $I$ is an identity matrix, and $B=(\sum^N_{j=1}a_j)I_{N+1}$.

The vector eigenfunctions of the linear system (\ref{lax}) can be calculated as
\begin{eqnarray}
\Psi_m=\begin{pmatrix} \Psi_{0m}+\tilde{\Psi}_{0m}  \\
\frac{\psi_{0}^{(1)}{\Psi}_{0m}}{(\beta_1+{\bm\chi}_m)}+\frac{\psi_{0}^{(1)}\tilde{\Psi}_{0m}(\bm\chi_m+\beta_1-1)}{(\beta_1+\bm\chi_m)^2}\\
\frac{\psi_{0}^{(2)}{\Psi}_{0m}}{(\beta_2+{\bm\chi}_m)} +\frac{\psi_{0}^{(2)}\tilde{\Psi}_{0m}(\bm\chi_m+\beta_2-1)}{(\beta_2+\bm\chi_m)^2} \\
\vdots\\
\frac{\psi_{0}^{(N)}{\Psi}_{0m}}{(\beta_N+{\bm\chi}_m)}+\frac{\psi_{0}^{(N)}\tilde{\Psi}_{0m}(\bm\chi_m+\beta_N-1)}{(\beta_N+\bm\chi_m)^2} \end{pmatrix},\label{psim}
\end{eqnarray}
where
\begin{eqnarray}
\Psi_{0m}&=&i(\bm x+\bm\chi_m\bm t+1)\tilde{\Psi}_{0m},~~~\\
\tilde{\Psi}_{0m}&=&\exp[i\bm\chi_m(\bm x+\bm\chi_m\bm t/2)].\label{vg1}
\end{eqnarray}
with $\bm x=x-x_{0m}$, and $\bm t=t-t_{0m}$~($m=1,...,M$). Here, the spectral parameter should be: $$\lambda_m=\bm\chi_m-\sum^N_{j=1}\frac{a_j}{\bm\chi_m+\beta_j},$$ with
$\bm\chi_m$ being the eigenvalue of the linear Lax pair system, which is given by
\begin{eqnarray}
1+\sum_{j=1}^{N}\frac{a_j^2}{(\bm{\chi}_m+\beta_j)^2}=0.\label{eqchi-aj}
\end{eqnarray}
When $a_j=a$, Eq. (\ref{eqchi-aj}) reduces to (\ref{eqchi}).

When $M=1$, we obtain the fundamental (first-order) RW solution by performing the following Darboux transformation:
\begin{eqnarray}
&&\psi^{(j)}[1]=\psi_{0}^{(j)}+(\lambda^*_1-\lambda_1)(P[1])_{j+1, 1},\\
&&P[1]=\frac{\Psi_1\Psi^\dag_1}{\Psi^\dag_1\Psi_1}.
\end{eqnarray}
Here, $\ast$ represents the complex conjugation, $\Psi_1$ is the special solution (\ref{psim}) as $\bm\chi_m=\bm\chi_1$.
$(P[1])_{j+1, 1}$ represent the elements of the matrix $P[1]$ in the first column, ($j+1$)th row.
In fact, the simplified form of the fundamental RW solution is given by Eq. (\ref{eqrw}).
Fundamental RWs with different eigenvalues ($\pm\bm{\chi}_{(2)}$ for $N=2$; $\pm\bm{\chi}_{(2)}$, $\bm{\chi}_{(1)}$ for $N=3$) are shown in Figs. \ref{fn2} and \ref{fn3}, respectively.

However, to obtain non-degenerate multi-RWs reported here, we perform the second step of the Darboux transformation.
We employ $\Psi_2$ [solution (\ref{psim}) as $\bm\chi_m=\bm\chi_2$] which is mapped to
\begin{eqnarray}
&&\Psi_2[1]=T [1]|_{\lambda=\lambda_2}\Psi_2, \nonumber\\
&&T[1]=I+\frac{\lambda^*_1-\lambda_1}{\lambda-\lambda^*_1}P[1].
\end{eqnarray}
Then, the second-order RW solution can be given by:
\begin{eqnarray}
&&\psi^{(j)}[2]=\psi^{(j)}[1]+(\lambda^*_2-\lambda_2)(P[2])_{j+1,1},\nonumber\\
&&P[2]=\frac{\Psi_2[1]\Psi^\dag_2[1]}{\Psi^\dag_2[1]\Psi_2[1]}.
\end{eqnarray}
Similarly, we can perform the $M$th-order Darboux transformation for the vector NLSEs (\ref{eqmanakov}) as follows:
\begin{eqnarray}
&&\Psi_M[M-1]=(T[M-1] ... T[2]T[1]\Psi_M)|_{\lambda=\lambda_M},\nonumber\\
&&T[M]=I+\frac{\lambda^*_M-\lambda_M}{\lambda-\lambda^*_M}P[M].
\end{eqnarray}
Then the $M$-th order solutions can be given by:
\begin{eqnarray}
&&\psi^{(j)}[M]=\psi^{(j)}[M-1]+(\lambda^*_M-\lambda_M)(P[M])_{j+1,1},~~~~\nonumber\\
&&P[M]=\frac{\Psi_M[M-1]\Psi^\dag_M[M-1]}{\Psi^\dag_M[M-1]\Psi_M[M-1]}.
\end{eqnarray}
For $N=2$, we obtained the non-degenerate second-order RWs ($M=2$) with two eigenvalues ($\pm\bm{\chi}_{(2)}$). The amplitude distributions for such RWs are shown in Figs. \ref{f3}(a) and \ref{f3}(b). For $N=3$, we presented the non-degenerate second-order RWs ($M=3$) with three eigenvalues ($\pm\bm{\chi}_{(2)}$, $\bm{\chi}_{(1)}$). Examples of such RWs are shown in Figs. \ref{f3}(c) and \ref{f3}(d).

\end{appendix}

\end{CJK}

\end{document}